\documentstyle[12pt,epsf]{article}

\global\arraycolsep=1pt
\begin{document}

\newcommand{\bx}{{\bf x}}
\newcommand{\by}{{\bf y}}
\newcommand{\tbx}{{\tilde \bx}}
\newcommand{\bu}{{\bar u}}
\newcommand{\bv}{{\bar v}}
\newcommand{\bz}{{\bf z}}
\newcommand{\bn}{{\bf n}}
\newcommand{\bm}{{\bf m}}

\newcommand{\ef}{e^{\varphi}}
\newcommand{\ew}{e^{2\varphi}}
\newcommand{\emf}{e^{-\varphi}}
\newcommand{\e}{e^{-2\varphi}}
\newcommand{\epr}{e^{-2\varphi(\bx')}}

\newcommand{\bj}{{\bar J}}
\newcommand{\tj}{{\tilde J}}

\newcommand{\w}{\wedge}
\newcommand{\pr}{\partial}

\newcommand{\al}{\alpha}
\newcommand{\bal}{{\bar \alpha}}
\newcommand{\be}{\beta}
\newcommand{\bbe}{{\bar \beta}}
\newcommand{\gm}{\gamma}
\newcommand{\dl} {\delta}
\newcommand{\bdl}{{\bar \dl}}
\newcommand{\ep}{\epsilon}
\newcommand{\bep}{{\bar \epsilon}}
\newcommand{\f}{\varphi}
\newcommand{\kp} {\kappa}

\newcommand{\hf} {{1 \over 2}}
{\baselineskip=14pt 
\rightline{
 \vbox{\hbox{YITP-96-55}
       \hbox{hep-th/9610187}
       \vskip 2mm
       \hbox{October 1996}       }}}

\vskip 10mm
\begin{center}
{\Large \bf Two-toroidal Lie Algebra as 
 Current Algebra of \\ Four-dimensional K\"ahler WZW Model}

\vskip 12mm
{\bf
Takeo Inami, \ Hiroaki Kanno${}^{\ast}$, \ Tatsuya Ueno \\
and \\ Chuan-Sheng Xiong  }
\\ 

\vskip 10mm
\it{Yukawa Institute for Theoretical Physics}\\
\it{Kyoto University, Kyoto 606-01, Japan}\\
\it{ and }\\
\it{Department of Mathematics, Faculty of Science${}^{\ast}$}\\
\it{Hiroshima University, Higashi-Hiroshima 739, Japan} 

\end{center}

\vskip 18mm
\begin{center}
{\large \bf ABSTRACT}
\end{center} 
We investigate the structure of an infinite-dimensional symmetry 
of the four-dimensional K\"ahler WZW model, which is a possible 
extension of the two-dimensional WZW model. 
We consider the $SL(2,R)$ group and, using the Gauss decomposition 
method, we derive a current algebra identified with a two-toroidal 
Lie algebra, a generalization of the affine Kac-Moody algebra. 
We also give an expression of the energy-momentum tensor in 
terms of currents and extra terms. 
\thispagestyle{empty}
\newpage

\section{Introduction}%
Many important physical phenomena in particle physics, such as 
chiral symmetry breaking and confinement, are of a non-perturbative 
nature, and it is difficult to solve quantum field theories (QFT's) 
describing these phenomena.
There is, however, a large class of 1+1D QFT's which share many of 
the important properties of 3+1D QFT's. 
Some such theories, e.g. the Gross-Neveu model and the sine-Gordon 
model, have been proven to be exactly solvable, and they have provided 
us with insight into the properties of their 3+1D counterparts.
\par

The exact solvability of 1+1D QFT relies on the existence of an 
infinite number of conservation laws associated with 
infinite-dimensional symmetry, e.g. an affine Kac-Moody (KM) algebra 
symmetry in the Wess-Zumino-Witten (WZW) model.
Continual attempts have been made toward constructing integrable
QFT's in space-times of more than two dimensions \cite{PJ},
but without much progress.
4D Self-dual Yang-Mills and self-dual gravity equations have been 
shown to be completely integrable. 
However, these theories are at the classical level, and in four
dimensions, no integrable QFT's based on Lagrangian have been 
constructed.
\par

Recently a non-linear sigma model (NLSM) in four dimensions 
with a Wess-Zumino like term has been proposed as a possible 
extension of the 2D WZW model \cite{NS},\cite{LMNS}.
It was first introduced by Donaldson \cite{D} (see also \cite{P}).
In \cite{NS}, this model, which we refer to as 4D K\"ahler WZW 
(KWZW) model, was studied in relation to the 5D K\"ahler Chern-Simons 
theory and was shown to have an infinite-dimensional symmetry. 
Later, it was shown that the model is solvable in some
algebraic sector \cite{LMNS} and that it is one-loop on-shell finite 
\cite{LMNS},\cite{K}.
Interestingly, the KWZW model appears as an effective field 
theory of the $N = 2$ string \cite{OV} and also as a chiral sector 
of the QCD scattering theory \cite{CCS}.
\par

In this paper, we investigate the structure of the
infinite-dimensional symmetry of the KWZW model in 4D flat space-time.
To construct the current algebra for the symmetry explicitly and 
to simplify the argument, we consider the group $SL(2,R)$. 
The extension to other Lie groups will be mentioned later. 
By using the method of the Gauss decomposition, we derive the current
algebra and identify it with a two-toroidal Lie algebra, which is a 
generalization of the affine KM algebra. 
We also give an expression for the energy-momentum tensor in 
terms of currents and extra terms, in the spirit of Sugawara's 
conjecture.

\section{The KWZW model and Gauss decomposition}%
The basic field in the KWZW model is a mapping $g(x)$ from 
a four-manifold $X_4$ to a Lie group $G$. 
The action for $g(x)$ is given as a generalization of the 
2D WZW model \cite{NS},\cite{LMNS}, 
\begin{equation}
S = -{i \over 4\pi} 
   \int_{X_4} \omega \w {\rm Tr}(g^{-1} \pr g \w g^{-1} {\bar \pr} g)
  +{i \over 12\pi} \int_{X_5} \omega \w {\rm Tr}(g^{-1} d g)^3 \ .  
                                                \label{eq: Act} 
\end{equation}
The two-form $\omega$ is a closed K\"ahler form on $X_4$, 
\begin{equation}
 \omega = {i \over 2} f^2_{\pi} \, h_{\al {\bar \be}} \, 
dz^{\al} \w d{\bar z}^{\be} \ ,
\end{equation}
where $h_{\al \bbe}$ $(\al, \bbe = 1,2)$ is a K\"ahler metric on
$X_4$ and $f_{\pi}$ is a coupling constant with the mass dimension $+1$.
The equation of motion reads,
\begin{equation}      
 {\bar \pr} ( \omega \w g^{-1} \pr g) = 0 \ ,      \label{eq: em1}
\end{equation}
or equivalently,
\begin{equation}
 \pr (\omega \w {\bar \pr} g \, g^{-1}) = 0 \ .    \label{eq: em2}
\end{equation}
These equations are known as the Yang equations on a K\"ahler manifold 
$X_4$ and are equivalent to the self-dual Yang-Mills (SDYM) equation 
in a particular gauge \cite{Y}. 
When we restrict $X_4$ to a hyper-K\"ahler manifold, the two-form 
$\omega$ represents a self-dual gravitational (SDG) instanton, since 
in four dimensions the hyper-K\"ahler condition is equivalent to the 
self-duality of the Riemann tensor. 
Then (\ref{eq: em1}) or (\ref{eq: em2}) describes a SDYM instanton  
coupled to a SDG instanton.
\par 

A comment is in order on the signature of the space-time $X_4$. 
The $(4,0)$ signature is favoured from the viewpoint of the SDYM 
equation, whereas the KWZW model arising from the $N=2$ string 
and that in the QCD prefer $(2,2)$ and $(3,1)$, respectively.
In this paper, we will not be concerned with the signature of 
$X_4$, assuming that any two signatures are connected to each other 
by analytic continuation.
\par

The integrable self-dual equation suggests the existence of an  
infinite-dimensional symmetry in the action (\ref{eq: Act}).
In fact, we can prove the following identity \cite{NS},\cite{LMNS}, 
which is an analogue of the Polyakov-Wiegmann formula,
\begin{equation}
 S[gh] = S[g] + S[h] - {i \over 2\pi} 
       \int_{X_4}\omega \w {\rm Tr}g^{-1}\pr g {\bar \pr}h h^{-1} \ . 
                                           \label{eq: PW}
\end{equation}
{}From this formula, we can easily see that the action is invariant 
under holomorphic right and anti-holomorphic left symmetries, 
$g \rightarrow h_L({\bar z}^1, {\bar z}^2) g h_R(z^1, z^2)$. 
Corresponding to the right (left) action symmetry, we have a conserved 
current $J$ ($\bj$),
\begin{equation}
J = {i \over \pi} \omega \w g^{-1} \pr g \ , 
\ \ \ \ \ \    
\bj = - {i \over \pi} \omega \w {\bar \pr} g \, g^{-1} \ .
                                           \label{eq: cc}
\end{equation}
\par 

It was known that the SDYM equation has another type of symmetry 
associated with an infinite number of non-local conserved currents
\cite{DC}, which is different from the symmetry of the KWZW model
associated with (\ref{eq: cc}).
The former symmetry leads to an affine KM algebra with no central 
extension.
On the other hand, as will see below, we do have a central extension
in the infinite-dimensional algebra for the KWZW model.
The symmetry in \cite{DC} is an analogue of that in the 2D principal
chiral model, whereas the symmetry in this paper is an analogue of 
that in the 2D WZW model.
\par

Although the KWZW model is apparently non-renormalizable by power 
counting, it has been shown that the model is one-loop on-shell
finite \cite{LMNS}. 
Conversely, the requirement of the one-loop finiteness of 
the NLSM with torsion leads to the KWZW model uniquely \cite{K}.
This remarkable property arises from the existence of 
torsion in group manifolds, represented by the Wess-Zumino like 
term, the second term in (\ref{eq: Act}). 
\par

\vskip 5mm

We consider the KWZW model with the $SL(2,R)$ group and use 
the Gauss decomposition method to express  $g(x)$ in the neighborhood 
of the unit element as follows,
\begin{equation}
g(x) =  e^{\chi(x) E_+} e^{\varphi(x) H} e^{\psi(x) E_-} =
\left[\begin{array}{cc} \ef + \chi \psi \emf \ &\ \chi \emf \\ 
\psi \emf \ &\ \emf \end{array} \right] \ ,                             
\end{equation}
where $\varphi(x)$, $\chi(x)$ and $\psi(x)$ are real scalar fields
and 
\begin{equation}
H =\left[\begin{array}{cc} 1 \ &\ 0 \\ 0 \ &\ -1 \end{array} \right] 
\ ,\ \ 
E_+ =\left[\begin{array}{cc} 0 \ &\ 1 \\ 0 \ &\ 0 \end{array} \right]
\ , \ \ 
E_- =\left[\begin{array}{cc} 0 \ &\ 0 \\ 1 \ &\ 0 \end{array} \right]
\ .
\end{equation}
We also use the Hermite basis of the $sl(2)$ algebra, $T^i = \hf \sigma^i$
($i=1,2,3$), where $\sigma^i$ are Pauli matrices.
Then we make use of the identity (\ref{eq: PW}) to re-express the
action (\ref{eq: Act}) as 
\begin{equation}
S =  -{i \over 2\pi} \int_{X_4} \omega \w [ \pr \f \w {\bar \pr} \f 
     + \pr \chi \w {\bar \pr} \psi \, \e] \ .
\end{equation}
\par

In this paper, we assume that the space-time $X_4$ is flat;
$h_{1 {\bar 1}} = h_{2 {\bar 2}} = \hf$ with the others zero, 
and use the notation $u = z^1,\ v = z^2$ 
$(\bu = {\bar z}^1,\ \bv = {\bar z}^2)$. 
Then, the action reads
\begin{equation}
S =  \kp \int d^4 z ( \pr_{\gm} \f \pr_{{\bar \gm}} \f 
    + \pr_{\gm} \chi \pr_{{\bar \gm}} \psi \,\e) \ ,
\ \ \ \ \ \kp = - {f^2_{\pi} \over 8\pi} \ .     \label{eq: act}
\end{equation}
The energy-momentum tensor is given by 
\begin{eqnarray}
T_{\al \be} &&= {\kp \over 2}(\pr_{\al} \f \pr_{\be} \f
              + \pr_{\al} \psi \pr_{\be} \chi \, \e) \ , 
\nonumber \ \ \ \ \ \ 
T_{{\bar \al} {\bar \be}} = {\kp \over 2}
               (\pr_{\bar \al} \f \pr_{\bar \be} \f
                + \pr_{\bar \al} \chi \pr_{\bar \be} \psi \, \e) \ ,
\\
T_{\al {\bar \be}} && = T_{{\bar \be} \al} 
= {\kp \over 2}
(\dl_{\al}^{\gm} \dl_{\bar \be}^{\bdl} 
 - h_{\al {\bar \be}} h^{\gm \bdl}) 
(\pr_{\gm} \f \pr_{\bdl} \f
    + \pr_{\gm} \chi \pr_{\bdl} \psi \, \e) \ ,    \label{eq: E-M}
\end{eqnarray}
with their conservation laws,
\begin{equation}
\pr_{\be} T_{\al {\bar \be}} + \pr_{\bar \be} T_{\al \be} = 0 \ , 
\ \ \ \ 
\pr_{\be} T_{{\bar \al} {\bar \be}} + \pr_{\bar \be} T_{{\bar \al}
\be} = 0 \ . 
\end{equation}
\par

The components of the current $J = J^0 H + J^- E_+ + J^+ E_- $ are 
given in terms of $\varphi(x)$, $\chi(x)$ and $\psi(x)$,
\begin{eqnarray}
J^0 &&= {i \over \pi} \omega \w (\pr \f + \psi \pr \chi \, \e) \ ,
\nonumber \\
J^+ &&= {i \over \pi} \omega \w 
        (\pr \psi - 2 \pr \f \psi - \psi^2 \pr \chi \, \e) \ ,
\ \ \ \ 
J^- = {i \over \pi} \omega \w \pr \chi \, \e  \ , 
\end{eqnarray}
whose $u$, $v$ components are defined from the conservation 
laws, 
${\bar \pr} J^a = h^{\al {\bar \be}} \pr_{\bar \be} J^a_{\al} d^4z =0$,
$(a=0,+,-)$, 
\begin{equation}
J^0_{\al} = \kp (\pr_{\al} \f + \psi \pr_{\al} \chi \, \e) \ ,
\ \ 
J^+_{\al} = \kp
(\pr_{\al} \psi - 2 \pr_{\al} \f \psi - \psi^2 \pr_{\al} \chi \, \e) 
\ , \ \ 
J^-_{\al} = \kp \pr_{\al} \chi \, \e  \ .  \label{eq: Jal}
\end{equation}
The currents $J^a$ correspond to the holomorphic 
right-action symmetry with parameters $\ep_a (z^1, z^2)$;  
$\dl_R g (x) = \ep_0 g H + \ep_+ g E_+ + \ep_- g E_- $. 
Accordingly $\varphi(x)$, $\chi(x)$ and $\psi(x)$ transform 
in the following way,
\begin{equation}
\dl \f = \ep_0 - \ep_+ \psi \ , \ \ \ \
\dl \psi = 2 \ep_0 \psi - \ep_+ \psi^2 + \ep_- \ , \ \ \ \
\dl \chi = \ep_+ e^{2 \f} \ .         \label{eq: Jt} 
\end{equation}
\par

Similarly, the components of the current $\bj$ are given by
\begin{eqnarray}
\bj^0 &&= - {i \over \pi} 
         \omega \w ({\bar \pr} \f + \chi {\bar \pr} \psi \, \e) \ , 
\nonumber \\
\bj^+ &&= - {i \over \pi} \omega \w {\bar \pr} \psi \, \e \ ,
\ \ \ \ 
\bj^- = - {i \over \pi} \omega \w 
({\bar \pr} \chi -2 {\bar \pr} \f \chi - \chi^2 {\bar \pr} \psi \, \e)
\ . 
\end{eqnarray}
The $\bu$, $\bv$ components of the $\bj^a$ are 
\begin{equation}
\bj^0_{\bbe} = \kp (\pr_{\bbe} \f + \chi \pr_{\bbe} \psi \, \e) \ ,
\ \ 
\bj^+_{\bbe} = \kp \pr_{\bbe} \psi \, \e \ ,
\ \
\bj^-_{\bbe} = \kp
(\pr_{\bbe} \chi -2 \pr_{\bbe} \f \chi - \chi^2 \pr_{\bbe} \psi \, \e) 
\ .    \label{eq: Jbbe}
\end{equation}
The currents $\bj^a$ correspond to the anti-holomorphic left-action 
symmetry with parameters $\bep_a ({\bar z}^1, {\bar z}^2)$; 
$\dl_L g (x) = \bep_0 H g  + \bep_+  E_+ g  + \bep_-  E_- g$.
Accordingly $\varphi(x)$, $\chi(x)$ and $\psi(x)$ transform as  
\begin{equation}
\dl\f = \bep_0 - \bep_- \chi \ , \ \ \ \
\dl\psi = \bep_-  e^{2\f}       \ , \ \ \ \ 
\dl\chi = 2 \bep_0 \chi + \bep_+ - \bep_- \chi^2 \ . \label{eq: bjt}
\end{equation}

\section{Two-toroidal Lie algebra}%
To obtain a current algebra in the KWZW model, we start with the 
classical Poisson bracket (P.B.) in the light-cone frame. 
We take $\bu$ as our time coordinate, while the 
space variables are denoted as $\bx = (u, v, \bv)$.
In this light-cone frame, the action (\ref{eq: Act}) or 
(\ref{eq: act}) is first order in time derivatives, that is, 
it is already in Hamiltonian form. 
Therefore, we can define the P.B. without introducing conjugate 
momenta of fields $\phi^i = (\varphi, \psi, \chi)$.
The formula applied to the 2D WZW model \cite{W}
can be easily extended to the present case as follows,  
\begin{equation}
\{ X(\bx), Y(\bx') \}  
= \int d^3 \tbx d^3 \tbx'  
\, \pr X(\bx) /\pr \phi^i(\tbx) \, F^{ij}(\tbx, \tbx') \,
\pr Y(\bx') /\pr\phi^j(\tbx') \ ,        \label{eq: PB}
\end{equation}
where $F^{ij}$ is the inverse of the matrix $F_{ij}$ defined from 
the variation of the action, 
$\dl S = \int d\bu \int d^3\bx d^3\bx' F_{ij}(\bx, \bx') \,  
\dl\phi^i (\bx) \pr_{\bu} \phi^j (\bx')$, 
\begin{equation}
F^{ij}(\bx,\bx') = {1 \over 2\kp}
\left[
\begin{array}{ccc}
-\hf &\ 0 \ & \chi(\bx) - \chi(\bx')  \\
     0       & 0 & - e^{2\f(\bx)}                  \\
-(\chi(\bx) - \chi(\bx')) & - e^{2\f(\bx')}
& (\chi(\bx) - \chi(\bx'))^2
\end{array}
\right]
\ep (u - u') \dl^2 (v - v') \ .
\end{equation}
The function $\ep (u - u')$ is the sign function,
and $\dl^2 (v - v') = \dl (v - v') \dl (\bv - \bv')$.
Henceforth, we take the values of the coordinates real 
by using an analytic continuation.
\par

Since $\bu$ is our time coordinate, from the conservation laws, 
$J^i_u$ are identified with the generators of the $J$-transformation 
(\ref{eq: Jt}), while $T_{uu}$, $T_{vu}$, $T_{\bv u}$ 
the generators for space translations and $T_{\bu u}$ the Hamiltonian 
density. 
If we take the $u$ as the time coordinate instead and define the
P.B. with respect to $u$, then $\bj^i_{\bu}$ become the generators of 
the $\bj$-transformation (\ref{eq: bjt}). 
\par

In the following, we present various P.B.'s involving the currents 
$J^i_u$.
Using (\ref{eq: PB}), we have fundamental P.B.'s for the components of
the group element $g(\bx)$,
\begin{equation}
\{g_{ij}(\bx), g_{kl}(\bx') \} = {1 \over 4\kp} 
( g_{ij}(\bx)g_{kl}(\bx') - 2 g_{kj}(\bx)g_{il}(\bx'))
\ep(u-u') \dl^2(v-v') \ . 
\end{equation}
P.B.'s of $g(x)$ and $J^a_u$ are given in the matrix form as
\begin{eqnarray}
&& \{g(\bx), J^-_u(\bx') \}  = g(\bx) E_- \, \dl^3(\bx-\bx') \ ,
\ \ \ \ \ 
\{g(\bx), J^+_u(\bx') \}  = g(\bx) E_+ \, \dl^3(\bx-\bx') \ ,
\nonumber \\
&& \{g(\bx), J^0_u(\bx') \}  = \hf g(\bx) H \, \dl^3(\bx-\bx') \ ,
                                               \label{eq: gj}
\end{eqnarray}
or, in the Hermite basis, 
$J^3 = J^0$, $J^1 = \hf(J^+ +J^-)$, $J^2 = {1 \over 2i} (J^+ - J^-)$, 
the above P.B.'s are expressed concisely as 
\begin{equation}
\{g(\bx), J^i_u(\bx') \} = g(\bx) T^i \, \dl^3(\bx-\bx') \ .
\end{equation}
{}From (\ref{eq: gj}), we can easily see that $J^a_u$ generate the 
transformation (\ref{eq: Jt}).
For later use, we introduce $G^{ij} = \hf {\rm Tr}T^i g^{-1} T^j g$, 
which satisfies the following identities:
$G^{il} \, G^{jk} \, \dl_{ij} = \dl_{lk}$,
$G^{li} \, G^{kj} \, \dl_{ij} = \dl_{lk}$.
The P.B.'s of the $G^{ij}$ and the currents $J^k_u$ are 
\begin{equation}
\{ G^{ij}(\bx), J^k_u(\bx') \} = -i \ep^{ikl} G^{lj}(\bx) 
                                    \dl^3 (\bx-\bx')        \ .
\end{equation}
\par

P.B.'s among $J^i_u$'s are in the Hermite basis,
\begin{equation}
\{J^i_u(\bx), J^j_u(\bx') \} =
              -i \ep^{ijk} J^k_u(\bx) \dl^3(\bx - \bx') 
              + {\kp \over 2} \dl^{ij} \pr_u \dl^3(\bx - \bx') \ ,
\end{equation}
and those among $J^i_v$ and $J^j_u$ are 
\begin{equation}
\{J^i_v(\bx), J^j_u (\bx') \} =
              -i \ep^{ijk} J^k_v(\bx) \dl^3(\bx - \bx') 
              + {\kp \over 2} \dl^{ij} \pr_v \dl^3(\bx - \bx') \ .
\end{equation}
P.B.'s of $J^i_u$ with $\bj^j_{\bv}$ and $\bj^j_{\bu}$ do not vanish,
unlike the case of the 2D WZW model;  
\begin{eqnarray}
\{\bj^j_{\bv}(\bx), J^i_u(\bx') \} &&= 
-{\kp \over 2} \pr_{\bv'} (G^{ij}(\bx') \, \dl^3(\bx - \bx')) \ , 
                                        \label{eq: bjv-ju}
\\
\{ \bj^j_{\bu}(\bx), J^i_u(\bx') \} &&= 
{\kp \over 4} \pr_{\bv'} ( G^{ij}(\bx') \ep(u-u') \pr_v \dl^2(v-v') )
\ .                                     \label{eq: bju-ju}
\end{eqnarray}
Note, however, that right-hand sides of (\ref{eq: bjv-ju}) and 
(\ref{eq: bju-ju}) are total derivatives in the coordinate $\bv'$ and 
hence can be dropped when we integrate the currents $J^i_u(\bx')$ with 
respect to $\bv'$.
\par

\vskip 5mm

To define our current algebra properly, we compactify space directions 
such that $u$, $v$ and $\bv$ take the values $I=[0, 2\pi]$.
The charge of the $J$-transformation (\ref{eq: Jt}) is then given by  
\begin{equation}
Q^i = \int^{2\pi}_0 du dv \ep_i(u,v) \tj^i \ ,
\ \ \ \ \
\tj^i = \int^{2\pi}_0 d\bv J^i_u(\bx) \ . 
\end{equation}
{}From the conservation laws, we see that $\tj^i$ are functions of 
$u$ and $v$ only; $\tj^i = \tj^i(u,v)$. 
We use the notation $\bz = (u,v)$. 
Then the P.B.'s for $\tj^i(\bz)$ are written as
\begin{equation}
\{ \tj^i(\bz), \tj^j(\bz') \}
= -i \ep^{ijk} \tj^k(\bz) \dl^2(\bz-\bz') 
+ \pi\kp \dl^{ij} \pr_u \dl^2 (\bz-\bz') \ .    \label{eq: tj}
\end{equation}
This is also noted in \cite{NS}.
We list other P.B.'s involving $\tj^i(\bz)$,
\begin{eqnarray}
&& \{ g(\bx), \tj^i(\bz') \} = g(\bx)T^i \dl^2(\bz-\bz')   \ ,
\\
&& \{ G^{ij}(\bx), \tj^k(\bz') \} = -i \ep^{ikl} G^{lj}(\bx) 
                                     \dl^2 (\bz-\bz')       \ ,
\\
&&\{J^i_v(\bx), \tj^j (\bz') \} =
              -i \ep^{ijk} J^k_v(\bx) \dl^2(\bz - \bz') 
              + {\kp \over 2} \dl^{ij} \pr_v \dl^2(\bz - \bz') \ , 
\\
&& \nonumber 
\\
&&\{\bj^i_{\bv}(\bx), \tj^j(\bz') \} = 0 \ ,
\ \ \ \ \ \ \ \ 
\{\bj^i_{\bu}(\bx), \tj^j(\bz') \}= 0 \ .      \label{eq: bjj}
\end{eqnarray}
The equations in (\ref{eq: bjj}) mean that the currents $\tj^j$ 
are decoupled from $\bj^i_{\bu}$ and $\bj^i_{\bv}$, which is 
reminiscent of the decoupling of $J_z$ and $\bj_{\bar z}$ sectors 
in the 2D WZW model. 
\par

The parameter functions $\ep_i(\bz)$ are defined on the torus and 
can be expanded as 
\begin{equation}
\ep_i(\bz) = \sum_{\bn} \ep^{\bn}_i \, e^{i\bn \cdot \bz} \ ,
                                               \label{eq: par}
\end{equation}
where $\bn = (n_1, n_2) \in {\bf Z}^2$ and $\bn \cdot \bz = 
n_1u + n_2v$. 
Correspondingly, the modes of the currents $\tj^i(\bz)$ are defined as 
\begin{equation}
Q^i = \sum_{\bn} \ep^{\bn}_i J^i_{\bn} \ ,
\ \ \ \ \ 
J^i_{\bn} = \int du dv e^{i\bn \cdot \bz} \tj^i(\bz)    \ .
\end{equation}
Then from (\ref{eq: tj}) we obtain
\begin{equation}
\{J^i_{\bn}, J^j_{\bm} \} = -i \ep^{ijk} J^k_{\bn+\bm} 
 + i \lambda_1 n_1 \dl^{ij}\dl_{\bn, -\bm} \ ,
\ \ \ \ \lambda_1 = {\pi^2 f^2_{\pi}\over 2} \ .    \label{eq: 2t}
\end{equation}
\par

We now present the mathematical interpretation of the P.B. algebra 
(\ref{eq: 2t}) obeyed by $J^i_{\bn}$. 
We first note that, by restricting $\bn$ to the subset $\bn =
(n_1,0)$, (\ref{eq: 2t}) is reduced to the familiar affine KM algebra 
$\hat{sl}(2)$.
For general $\bn$, (\ref{eq: 2t}) defines an infinite-dimensional 
(i.e.\,$\bn \in {\bf Z}^2$) Lie algebra with central extension.
A few mathematicians have recently begun the study of the
representation theory of this class of algebra \cite{tta}.
In their terminology, (\ref{eq: 2t}) is called the two-toroidal 
Lie algebra $sl(2)_{tor}$ because of the double Fourier mode expansion 
(\ref{eq: par}). 
It is interesting that two-toroidal Lie algebra is obtained from 
the 4D KWZW model in the same fashion as the affine KM (one-toroidal)
algebra from the 2D WZW model.
\par

The root system of $\hat{sl}(2)$ is two-dimensional, being 
infinite in one direction. 
For comparison, we have tentatively chosen a root system for 
the $sl(2)_{tor}$ in close analogy with $\hat{sl}(2)$.
It is then three-dimensional, being infinite in two directions. 
The Cartan matrices of affine KM algebras have vanishing determinants, 
while those of hyperbolic KM algebras have negative determinants.
It is yet to be studied whether one can assign Cartan matrices to
general n-toroidal Lie algebras. 
For the $sl(2)_{tor}$ root system we have chosen, the  
``Cartan matrix'' $K$ is 
\begin{equation}
K=
\left[
\begin{array}{rrr}
2 \ & -2 \ & -2 \ \\ -2 \ & 2 \ & 2 \ \\ -2 \ & 2 \ & 2 \ 
\end{array}
\right] \ .
\end{equation}
The matrix $K$ has the features not shared by finite-dimensional and 
affine Lie algebras: its determinant has a double zero, and one of its 
off-diagonal elements is $+2$, which is shown by a dotted double line 
in the corresponding Dynkin diagram in Fig.1.
\par
\begin{figure}[b]
\epsfxsize= 35 mm
\begin{center}
\leavevmode
\epsfbox{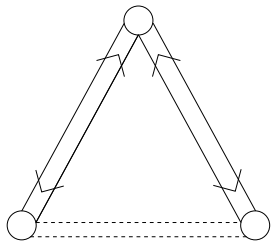}
\end{center}
\caption{The Dynkin diagram for $sl(2)_{tor}$. Lower two points are
simple roots for two null directions while the upper point is the
original $sl(2)$ simple root.}
\end{figure}

The representation content of affine KM algebras can be understood
based on highest-weight (HW) representations.
There is no obvious extension of HW representations to two-toroidal
Lie algebras because of the two-dimensionality of the infinite 
directions of their root systems.
This makes it difficult to find a criterion for good
representations.
\par

Let us focus on the central term in (\ref{eq: 2t}). 
There appears only one center proportional to $\lambda_1 n_1$, 
whereas a more general central term should consist of two terms, 
$\lambda_1 n_1 + \lambda_2 n_2$, by adding the new central charge 
$\lambda_2$.
The $\lambda_1$ has to be quantized to ensure that the measure 
$\exp iS$ in the path-integral is well-defined \cite{LMNS}, 
as in the 2D case.
See also \cite{NS}.
It is not clear how the quantization condition is related to  
the representation theory of the two-toroidal Lie algebra.
\par

The two-toroidal Lie algebra is completed by constructing
two kinds of derivations, $d_1$ and $d_2$, corresponding to 
$(n_1,n_2)$.
We will see later that $d_1$ and $d_2$ are given by the components of 
the energy-momentum tensor $T_{uu}$ and $T_{vu}$, respectively, 
integrated over the space coordinates $\bx$.

\section{Sugawara-like construction}%
We now look into the possibility of the KWZW model being a good field 
theory realizing Sugawara's conjecture of the theory of currents
\cite{S}.
{}From (\ref{eq: E-M}), (\ref{eq: Jal}), (\ref{eq: Jbbe}), we obtain 
the following expression for the energy-momentum tensor,
\begin{eqnarray}
T_{\al \be}  
&& = {1 \over 2\kp} (J^0_{\al} J^0_{\be} + \psi J^0_{\al} J^-_{\be}
         - \psi J^-_{\al} J^0_{\be} + J^+_{\al}J^-_{\be}) 
= {1 \over 2\kp} J^i_{\be} J^j_{\al} M_{ij} \ , 
\nonumber \\
T_{\bal \bbe}
&& = {1 \over 2\kp} (\bj^0_{\bal} \bj^0_{\bbe} 
+ \chi \bj^0_{\bal} \bj^+_{\bbe} - \chi \bj^+_{\bal} \bj^0_{\bbe}
+ \bj^+_{\bbe} \bj^-_{\bal} )
= {1 \over 2\kp} \bj^i_{\bal} \bj^j_{\bbe} {\bar M}_{ij} \ ,
\end{eqnarray}
where 
\begin{equation}
\begin{array}{ll}
M_{ij} = \dl_{ij} + b_{ij} 
& \ \ \ \ 
{\bar M}_{ij} = {}^TG^{ik}(\dl_{kl} + b_{kl})G^{lj}
              = \dl_{ij} + ({}^TG b G)_{ij}
\\ \ \ \ \ \ \
= I + 
\left[
\begin{array}{ccc}
0 \ & i \ & \psi \ \\ -i \ & 0 \ & -i \psi \ \\ - \psi \ & i \psi \ &
0 \ 
\end{array}
\right] \ ,
& \ \ \ \ \ \ \ \ \ \ \ \ \ \ \ \ \ \ \ \ \ \ \ \ \ \ \ \ \ \
  \ \ \ \ \ \ \,
= I + 
\left[
\begin{array}{ccc}
0 \ & i \ & -\chi \ \\ -i \ & 0 \ & -i\chi \ \\ \chi \ & i \chi \ & 
0 \   
\end{array}
\right] 
\end{array}
\ . 
\end{equation}
$T_{\al \bbe}$ reads
\begin{eqnarray}
T_{\al {\bar \be}} = T_{{\bar \be} \al}
&&= - {1 \over 2\kp} (\dl_{\al}^{\gm}\dl_{\bbe}^{\bdl} 
- h_{\al \bbe} h^{\gm \bdl})
(J^0_{\gm} \bj^0_{\bdl} - \chi J^0_{\gm} \bj^+_{\bdl}
- \psi J^-_{\gm} \bj^0_{\bdl} 
+ J^-_{\gm} \bj^+_{\bdl}(\psi\chi + \ew)) 
\nonumber \\
&&= - {1 \over 2\kp} (\dl_{\al}^{\gm}\dl_{\bbe}^{\bdl} 
- h_{\al \bbe} h^{\gm \bdl}) J^i_{\gm} \bj^j_{\bdl} N_{ij} \ ,
\end{eqnarray}
where 
\begin{equation}
N_{ij} = (\dl_{ik} + b_{ik}) G^{kj} = 
\left[
\begin{array}{ccc}
\psi \chi + \ew \ & i(\psi \chi + \ew) \ & - \psi \ \\
-i(\psi \chi + \ew) \ & \psi \chi + \ew \ & i \psi \ \\
- \chi \ & -i \chi \ & 1 \ 
\end{array}
\right]
\ .
\end{equation}
Note that $M_{ij}$, ${\bar M}_{ij}$ and $N_{ij}$ contain extra terms 
which consist of $G^{ij}$ and $b_{ij}$. 
The term $b_{ij}$ corresponds to the torsion potential, whose curl 
is the torsion of the $SL(2,R)$ group manifold.
This is different from the case of the 2D WZW model, where 
the torsion potential does not appear in the energy-momentum tensor. 
Thus, all components of the energy-momentum tensor are constructed 
from the set $(J^i_{\al}, \bj^j_{\bbe}, G^{ij}, b_{ij})$.
Note that $T_{uu}$, $T_{vv}$, $T_{\bu \bu}$ and  
$T_{\bv \bv}$ are described by currents only since $b_{ij}$ is 
anti-symmetric.  
\par

P.B.'s of the currents $J^k_u$ and the generators for space 
translations, $T_{uu}$, $T_{vu}$, $T_{\bv u}$ are given as
\begin{eqnarray}
\{T_{uu}(\bx), J^k_u (\bx')\} =&& \hf J^k_u(\bx)\pr_u \dl^3(\bx-\bx')
\ , \nonumber \\
\{T_{vu}(\bx), J^k_u(\bx')\}
= && \hf J^k_u(\bx) \pr_v \dl^3(\bx-\bx') 
+ {1 \over 4} \pr_u((J^k_v(\bx) + b_{kj}J^j_v(\bx)) \dl^3(\bx-\bx'))
\nonumber \\ 
&& \ \ \ \ \ \ \ \ \ \ \ \ \ \ \ \ \ \ \ \ \ \ \ \ \ \ \
- {1 \over 4} \pr_v((J^k_u(\bx) + b_{kj}J^j_u(\bx))
\dl^3(\bx-\bx')) \ ,
\\
\{T_{\bv u}(\bx), J^k_u(\bx') \}
= && \hf J^k_u(\bx) \pr_{\bv} \dl^3(\bx-\bx') 
+ {1 \over 4} \pr_u((J^k_{\bv}(\bx) + b_{kj}J^j_{\bv}(\bx))
\dl^3(\bx-\bx'))
\nonumber \\ 
&& \ \ \ \ \ \ \ \ \ \ \ \ \ \ \ \ \ \ \ \ \ \ \ \ \ \ \ 
- {1 \over 4} 
\pr_{\bv}((J^k_u(\bx) + b_{kj}J^j_u(\bx)) \dl^3(\bx-\bx')) 
\nonumber \ .
\end{eqnarray}
The P.B.'s of the $J^k_u$ and the Hamiltonian density $T_{\bu u}$ are 
\begin{eqnarray}
\{ T_{\bu u}(\bx), J^k_u(\bx') \}
=&& - \hf J^k_v (\bx) \pr_{\bv} \dl^3(\bx-\bx')
    - {1 \over 4} 
\pr_v ((J^k_{\bv} + b_{kj}J^j_{\bv}) \dl^3(\bx-\bx'))
\nonumber \\ 
&& \ \ \ \ \ \ \ \ \ \ \ \ \ \ \ \ \ \ \ \ \ \ \ \ \ \ \ \ \ 
 + {1 \over 4} \pr_{\bv}((J^k_v + b_{kj}J^j_v ) \dl^3(\bx-\bx')) 
\ . 
\end{eqnarray}
As promised above, $T_{uu}$ and $T_{vu}$ play the role of the
derivations $d_1$ and $d_2$, respectively.

\section{Discussion}%
An interesting physical application of the KWZW model is the 
computation of QCD multi-gluon amplitudes with all the same helicity, 
called maximally helicity violating (MHV) amplitudes \cite{CCS}. 
This arises from the fact that the equation of motion of the KWZW model 
coincides with the SDYM equation.
The latter equation has been pointed out to be related 
to a calculation technique of the MHV amplitudes \cite{B}. 
These amplitudes vanish at tree level and take a very 
simple form at one-loop. 
These simple results are believed to be due to the infinite-dimensional 
symmetry of the SDYM equation.
Therefore, it is intriguing to reproduce these results from the
viewpoint of the representation theory of two-toroidal current 
algebras. 
A relation between some MHV amplitudes and the $k=1$ KM
algebra on $CP^1$ was already given by Nair \cite{N}.
\par

It was suggested that the KWZW model can be obtained from an effective 
action of 4D chiral fermions coupled to the $G \times U(1)$ 
background gauge field \cite{LMNS2}. 
We wish to point out that the field strength of the additional $U(1)$ 
factor, the K\"ahler potential, can be identified with the K\"ahler
form $\omega$ in the KWZW model.
In this case, the K\"ahler potential becomes dynamical since  
its kinematic term is also included in the effective action, 
which leads to the Plebanski equation for self-dual gravity \cite{Pl}. 
The same action also appears as a space-time effective action in 
the $N=2$ string \cite{OV}.
The quantum integrability of the string theory suggests that the current
algebra and the classical integrability we have found in this paper 
survive in the quantum theory. 
\par

In this paper, we have constructed the two-toroidal Lie algebra in 
the KWZW model in the case of the $SL(2,R)$ group.
The extension to the general $SL(N,R)$ case is straightforward.
The fundamental P.B.'s for the group element $g(\bx) \in SL(N,R)$ are  
\begin{equation}
\{g_{ij}(\bx), g_{kl}(\bx') \} = {1 \over 2N\kp} 
( g_{ij}(\bx)g_{kl}(\bx') - N g_{kj}(\bx)g_{il}(\bx'))
\ep(u-u') \dl^2(v-v') \ ,
\end{equation}
and the $sl(N)_{tor}$ algebra is derived from them.
Note that the algebra (\ref{eq: 2t}) can be obtained also for cases 
when the Gauss decomposition is impossible, e.g. $G = SU(N)$.
However, the treatment becomes more complicated in such cases.
It appears that the two-toroidal algebra symmetry is responsible 
for the one-loop finiteness of the model. 
It is of interest to examine whether the same symmetry also renders
the theory finite at all higher loops.
\par

The NLSM in six dimensions is more badly ultra-violet divergent. 
One may add a K\"ahler Wess-Zumino term induced from the theory 
of 6D chiral fermions with the $G \times U(1)$ gauge field.
It is interesting to see whether the NLSM with the additional 
term becomes one-loop finite and then whether a three-toroidal 
Lie algebra associated with an infinite-dimensional symmetry 
emerges.
\par


\vskip 20mm

We would like to thank Kiyoshi Higashijima, Minoru Wakimoto and 
Yoshihisa Saito for useful discussions.
We are grateful to Patrick Dorey for a careful reading of the 
manuscript.
This work is supported partially by Grant in Aid of the Ministry of
Education, Science and Culture. 
T.U. is supported by the Japan Society for the Promotion of Science, 
No.\,6293.
C.X. is supported by the COE (Center of Excellence) researchers program
of the Ministry of Education, Science and Culture.

\vskip 20mm


\end{document}